\documentstyle [12pt,epsf] {article}

\newcommand{\sect}[1]{ \section{#1} \setcounter{equation}{0} }

\newcommand{\req}[1]{(\ref{#1})}
\newcommand{\nwc} {\newcommand}
\nwc{\hyp} {\hyphenation}
\nwc{\Z}{\ZZ}

\def\bfone{\relax{\rm 1\kern-.35em 1}}
\def\inbar{\vrule height1.5ex width.4pt depth0pt}
\def\IC{\relax\,\hbox{$\inbar\kern-.3em{\mss C}$}}
\def\ID{\relax{\rm I\kern-.18em D}}
\def\IF{\relax{\rm I\kern-.18em F}}
\def\IH{\relax{\rm I\kern-.18em H}}
\def\II{\relax{\rm I\kern-.17em I}}
\def\IN{\relax{\rm I\kern-.18em N}}
\def\IP{\relax{\rm I\kern-.18em P}}
\def\IQ{\relax\,\hbox{$\inbar\kern-.3em{\rm Q}$}}
\def\IR{\relax{\rm I\kern-.18em R}}
\def\ZZ{\relax{\hbox{\mss Z\kern-.42em Z}}}
\font\cmss=cmss10 \font\cmsss=cmss10 at 7pt
\def\ZZ{\relax\ifmmode\mathchoice
{\hbox{\cmss Z\kern-.4em Z}}{\hbox{\cmss Z\kern-.4em Z}}
{\lower.9pt\hbox{\cmsss Z\kern-.4em Z}}
{\lower1.2pt\hbox{\cmsss Z\kern-.4em Z}}\else{\cmss Z\kern-.4em
Z}\fi}

\nwc{\be}  {\begin{equation}}
\nwc{\ee}  {\end{equation}}
\nwc{\ba}  {\begin{array}}
\nwc{\ea}  {\end{array}}
\nwc{\bdm} {\begin{displaymath}}
\nwc{\edm} {\end{displaymath}}
\nwc{\bea} {\be\ba{lcl}}
\nwc{\eea} {\ea\ee}
\nwc{\bda} {\bdm\ba{lcl}}
\nwc{\eda} {\ea\edm}
\nwc{\bc}  {\begin{center}}
\nwc{\ec}  {\end{center}}
\nwc{\ds}  {\displaystyle}
\nwc{\bmat}{\left(\ba}
\nwc{\emat}{\ea\right)}
\nwc{\nn} {\nonumber}
\nwc{\nnn} {\nonumber \vspace{.2cm} \\ }
\nwc{\ra}{\rightarrow}
\nwc{\lra}{\longrightarrow}
\nwc{\p} {\partial}

\nwc{\ep} {\epsilon}
\nwc{\de} {\delta}
\nwc{\Th} {\Theta}
\nwc{\th} {\theta}
\nwc{\al} {\alpha}
\nwc{\si} {\sigma}
\nwc{\Si} {\Sigma}
\nwc{\om} {\omega}
\nwc{\Om} {\Omega}
\nwc{\Ga} {\Gamma}
\nwc{\ga} {\gamma}
\nwc{\bet} {\beta}
\nwc{\La} {\Lambda}
\nwc{\la} {\lambda}
\nwc{\Sc}  {{\cal S}}
\nwc{\Rc}  {{\cal R}}
\nwc{\Dc}  {{\cal D}}
\nwc{\Oc}  {{\cal O}}
\nwc{\Cc}  {{\cal C}}
\nwc{\gc}  {{\cal g}}
\nwc{\Pc}  {{\cal P}}
\nwc{\Mc}  {{\cal M}}
\nwc{\Ec}  {{\cal E}}
\nwc{\Fc}  {{\cal F}}
\nwc{\Hc}  {{\cal H}}
\nwc{\Kc}  {{\cal K}}
\nwc{\Wc}  {{\cal W}}

\nwc{\Xc}  {{\cal X}}
\nwc{\Gc}  {{\cal G}}
\nwc{\Zc}  {{\cal Z}}
\nwc{\Nc}  {{\cal N}}

\nwc{\xc}  {{\cal x}}
\nwc{\Ac}  {{\cal A}}
\nwc{\Bc}  {{\cal B}}
\nwc{\Uc} {{\cal U}}
\nwc{\Vc} {{\cal V}}
\nwc{\Lc} {{\cal L}}
\nwc{\Qc} {{\cal Q}}

\nwc{\lng} {\langle}
\nwc{\rng} {\rangle}

\nwc{\lf} {\left}
\nwc{\ri} {\right}

\nwc{\diag} {{\rm diag}}
\nwc{\inv}  {{\rm inv}}
\nwc{\mod}  {{\ \rm mod\ }}
\nwc{\dete}  {{\rm det}}
\nwc{\tr}  {{\rm tr}}
\nwc{\im}  {{\rm Im}}
\nwc{\re}  {{\rm Re}}

\nwc{\h} {\frac{1}{2}}
\nwc{\fc} {\frac}

\def\KK{\relax{\rm I\kern-.18em K}}
\def\RR{\relax{\rm I\kern-.18em R}}
\def\NN{\relax{\rm I\kern-.18em N}}
\def\PP{\relax{\rm I\kern-.18em P}}

\def\zz{\relax{\sf Z\kern-.3em Z}}
\def\ZZ{\relax{\sf Z\kern-.4em Z}}
\def\ZZZ{{\relax{\sf Z}\kern -.5em Z}}
\def\ZZZ{Z\kern -0.37em Z}
\def\QQ{{\rm \kern .25em
             \vrule height1.4ex depth-.12ex width.06em\kern-.31em Q}}
\def\CC{{\rm \kern .25em
             \vrule height1.4ex depth-.12ex width.06em\kern-.31em C}}
\nwc{\F}{\ _2F_1}

\if@twoside\oddsidemargin25mm
\else\oddsidemargin25mm\evensidemargin25mm\marginparwidth25mm\fi%
\begin{document}
\begin{titlepage}
{\sf
\begin{flushright}
{CERN--TH/96--260}\\
{NEIP--96/006}\\
{hep--th/9609130}\\
{September 1996}
\end{flushright}}
\vfill
\vspace{-1cm}
\begin{center}
{\large \bf Periods,  Coupling Constants and Modular Functions\\[5mm]
in N=2\ \ SU(2) SYM with Massive Matter$^{\mbox{\boldmath $\ast$}}$}
\vskip 1.2cm
{\sc A. Brandhuber$^{1,2}$} {\ \small \ and}
{\ \ \sc S. Stieberger$^{2}$}\\
\vskip 1.5cm
{\em $^1$CERN -- Theory Division} \\
{\em CH--1211 Gen\`eve 23, SWITZERLAND}
\vskip 1cm
{and}\\
\vskip .6cm
{\em $^2$Institut de Physique Th\'eorique}\\
{\em Universit\'e de Neuch\^atel}\\
{\em CH--2000 Neuch\^atel, SWITZERLAND}
\end{center}
\vfill

\thispagestyle{empty}

\begin{abstract}
We determine the mass dependence of the coupling constant for
N=2 SYM with $N_f=1,2,3$ and $4$ flavours.
All these cases can be unified in one analytic expression, given
by a Schwarzian triangle function.
Moreover we work out the connection to modular functions which
enables us to give explicit formulas for the periods. Using the form
of the $J$--functions we are able to determine in an elegant way 
the couplings and monodromies at the superconformal points.
\end{abstract}

\vskip 5mm \vskip0.5cm
\hrule width 5.cm \vskip 1.mm
{\small\small  $^\ast$ Supported by the
Swiss National Science Foundation and the EEC under contracts\\
SC1--CT92--0789 and European Comission TMR programme ERBFMRX--CT96--0045.\\
Email: Andreas.Brandhuber@cern.ch and stieberg@surya11.cern.ch}
\end{titlepage}

\sect{Introduction}

Initiated by the seminal work of Seiberg and Witten \cite{sw1,sw2}
N=2 supersymmetric gauge theories
have received a lot of interest during the last two years.
The moduli space of the Coulomb branch characterized by the scalar
fields of N=2 vector multiplets and the masses of the matter fields 

receive
quantum corrections which are fully under control. The Wilsonian
effective action is completely determined in terms of certain elliptic
curves. Although these theories can be treated analytically they
have interesting strong coupling behaviour such as confinement,
chiral
symmetry breaking. Moreover, at certain points in the moduli space,  
they
provide examples of non--trivial interacting N=2 superconformal
theories in four dimensions \cite{ad,apsw}.
The objects of interest in the exact solution are the period integrals
over homology cycles on these elliptic curves which can alternatively
be determined using Picard--Fuchs equations. Up to now this has been
accomplished for
the theories with $SU(2)$ gauge group with massless matter  
\cite{ito,bf1,bf2}
and higher gauge groups \cite{klt}. More recently, it has been
generalized to the massive cases of the $SU(2)$ theories
\cite{ohta}.

The purpose of the present article is to determine the periods for
the $SU(2)$ theories for different numbers $N_f$ of matter fields
using a unique
differential equation. We find that the coupling constant is given
in terms of a Schwarzian triangle function. This function depends
only on the $J$--invariant of the elliptic curve and thus unifies all
cases -- both massless and massive, $N_f = 1,\ldots, 4$. This is shown
in section 2. 

In section 3 we work
out the relation between periods and
certain modular functions of $SL(2,\Z)$ or subgroups thereof.
For several values of $N_f$ we calculate the discriminant,
the modulus $u$ and finally the periods as modular functions of
the coupling constant $\tau$.
We demonstrate the unifying power of the $J$ invariant as a
function of $u$, the masses and the scale $\La$. That allows us
in section 4 to
read off the monodromies around the singularities
in the moduli space and to obtain the effective coupling constants
and the monodromies at the superconformal points in a very simple
fashion.

\sect{The coupling constant $\tau$}

Let us consider the quartic curve
$y^2=ax^4+4 b x^3+6c x^2+4dx+e$ written in {\em Weierstrass form}

\be
y^2=4x^3-g_2 x-g_3\ ,
\ee
with

\bea
\ds{g_2}&=&\ds{ae-4 b d+3 c^2}\nnn
\ds{g_3}&=&\ds{ace+2bcd-ad^2-b^2e-c^3\ .}
\eea
This curve corresponds to the one--parameter family of curves
embedded in $CP^2$

\be\label{LG}
X_s: x_1^3+x_2^3+x_3^3-s x_1 x_2 x_3=0\ ,
\ee
with $g_2=3s(8+s^3)$ and $g_3=8+20s^3-s^6$ \cite{klry}.
In particular, this curve describes all cases $N_f=0,1,2,3,4$
of \cite{sw1,sw2}.
The variation of the period of the holomorphic one--form
$\fc{dx}{y}$ along a homology cycle $\Ga$

\be\label{integral}
\tilde \om_\Ga=\oint_\Ga \fc{dx}{y}
\ee
is described by the second--order ODE \cite{fricke}

\be
\fc{d^2\Om_\Ga}{dJ^2}+\fc{1}{J}
\fc{d\Om_\Ga}{dJ}+\fc{31J-4}{144J^2(1-J)^2}
\Om_\Ga=0\ ,
\label{unif}
\ee
with the $J$ function defined as

\be\label{defJ}
J = \frac{g_2^3}{\Delta} = \frac{g_2^3}{g_2^3 - 27 g_3^2}
\ee
and the normalized periods

\be
\tilde \om_\Ga=\sqrt{\fc{g_2}{g_3}} \Om_\Ga\ .
\ee
The solutions $\Om_\Ga(J)$ of
this equation arise from the general case of
the hypergeometric differential equation
corresponding to the {\em Riemann P-function}  \cite{forsyth}

\be\label{riemann}
P(J)=\lf[\ba{cccc} 0&\infty&1&\ \\[5mm]
              -\fc{1}{6}&0&\fc{1}{4}&J \\[5mm]
              \fc{1}{6}& 0&\fc{3}{4}&\  \ea\ri]\equiv
(1-J)^{\fc{1}{4}}J^{-\fc{1}{6}} \lf[\ba{cccc} 0&\infty&1&\ \\[5mm]
0&\fc{1}{12}& 0&J \\[5mm]
\fc{1}{3}& \fc{1}{12} &\fc{1}{2}&\  \ea\ri]\ .
\ee
Therefore two linear independent solutions of \req{unif}, appropriate for
the regime $|J|>1,\ |arg(1-J)|<\pi$for $|J|>1$, read e.g.

\bea\label{solutions}
\ds{\Om_0(J)}&=&\ds{2\pi i(1-J)^{\fc{1}{4}}\
J^{-\fc{1}{4}}\F\lf[\fc{1}{12},
\fc{5}{12},1,\fc{1}{J}\ri]\ ,}\nnn
\ds{\Om_1(J)}&=&\ds{-(\ln J+3\ln 12)\ \Om_0(J)-w_1(J)}
\eea
with:

\bea
\ds{w_1(J)}&=&\ds{(1-J)^{\fc{1}{4}}\ J^{-\fc{1}{4}}\sum_{n=1}^\infty
\fc{\lf(\fc{1}{12}\ri)_n\lf(\fc{5}{12}\ri)_n}{(n!)^2}h_nJ^{-n}}\nnn
\ds{h_n}&=&\ds{2\psi(n+1)-\psi\lf(\fc{1}{12}+n\ri)
-\psi\lf(\fc{5}{12}+n\ri)+\psi\lf(\fc{1}{12}\ri)+\psi\lf(\fc{5}{12}\ri)-
2\psi(1)\ .}
\eea
Here $(a)_n\equiv \Ga(a+n)/\Ga(n)$ is the Pochhammer symbol.
Of course, there may be constructed many other pairs of independet
solutions both by taking different linear combinations and
expanding at the other singular points $J=0,1$. 
The specific choice \req{solutions} will become clear in a moment and
we also will have to say more about that at the end of this section.

Since the zeros of the discriminant $\Delta(u_i)=g_2^3-27g_3^2=0$
correspond to $J(u_i)=\infty$,
from the $J\ra\infty$ limit we can extract
the strong coupling behaviour around the points $u_i$.
But we also have $J\ra \infty$ for $u\ra \infty$.
Therefore a property of the solutions \req{solutions} is that
they have the appropriate form for both the weak--coupling
and the strong--coupling expansions. In particular we obtain:

\bea\label{limits}
\tilde \om_0&\lra&\ds{(-1)^{\fc{1}{4}}\sqrt{\fc{g_2(u,m)}{g_3(u,m)}}\
,}\nnn
\tilde \om_1&\lra&\ds{-(-1)^{\fc{1}{4}}\sqrt{\fc{g_2(u,m)}{g_3(u,m)}}
\ (\ln J+3 \ln 12)\ .}
\eea
The factor $\sqrt{\fc{g_2}{g_3}}$ is the leading piece and it is that
part
which is responsible for the different leading behaviours at $u\ra
u_i$
and $u\ra\infty$.
Let us present two examples\footnote{While this paper was being typed
some related work appeared
in \cite{ms}. Here the integral \req{integral} was performed explicitly
for the three cases $N_f=1,2,3$, however in a separate way. Therefore,
neither the unifying r\^ole of $J$ nor the connection to elliptic
functions could be seen.}:

\begin{itemize}

\item \underline{$N_f=0:$}

For the weak--coupling patch $u \sim \infty :$

\bea
\ds{\fc{da_D}{du}}&=&\ds{\fc{1}{12\pi}\sqrt{\fc{g_2(u,m)}{g_3(u,m)}}
\Om_1(J)\ ,}\nnn
\ds{\fc{da}{du}}&=&\ds{\fc{(-1)^{\fc{1}{4}}}{24\pi}
\sqrt{\fc{g_2(u,m)}{g_3(u,m)}}  \Om_0(J)\ ,}
\eea
and  for the strong--coupling regime $u\sim \La^2$:

\bea\label{mono0}
\ds{\fc{da_D}{du}}&=&\ds{\fc{1}{12\pi}\sqrt{\fc{g_2(u,m)}{g_3(u,m)}}
\Om_0(J)\ ,}\nnn
\ds{\fc{da}{du}}&=&\ds{\fc{(-1)^{-\fc{1}{4}}}{24\pi}\sqrt{\fc{g_2(u,m)}
{g_3(u,m)}} \Om_1(J)\ .}
\eea

\item \underline{$N_f=1:$}

For the weak--coupling patch  $u \sim \infty :$

\bea
\ds{\fc{da_D}{du}}&=&\ds{-\fc{i}{24\pi}\sqrt{\fc{g_2(u,m)}{g_3(u,m)}}
\Om_1(J)\ ,}
\nnn
\ds{\fc{da}{du}}&=&\ds{-\fc{i}{24\pi}\sqrt{\fc{g_2(u,m)}{g_3(u,m)}}\Om_0(J)\ 
  ,}
\eea
and  for the strong--coupling regime $u\sim u_1$:

\bea\label{mono1}
\ds{\fc{da_D}{du}}&=&\ds{\fc{i}{24\pi}\sqrt{\fc{g_2(u,m)}{g_3(u,m)}}
\Om_0(J)\ ,}\nnn
\ds{\fc{da}{du}}&=&\ds{\fc{i}{24\pi}\sqrt{\fc{g_2(u,m)}
{g_3(u,m)}} \Om_1(J)\ . }
\eea

\end{itemize}
Indeed, using several hypergeometric identities \cite{erdelyi1},
known as quadratic and cubic identities,
and the explicit expressions for the $J$--functions \req{defJ}

\bea\label{Js}
1728J&=&\ds{\fc{(3\La^4+u^2)^3}{\Delta_0(u)}\ \ ,\ \ N_f=0\
,}\nnn
1728J&=&\ds{\fc{256(3\La_1^3m-4u^2)^3}{2^{14}\Delta_1(u,m)}
\ \ ,\ \ N_f=1\ ,}\nnn
1728J&=&\ds{\fc{(3\La_2^4 - 48\La_2^2m^2 +
64u^2)^3}{2^{18}\Delta_2(u,m)}
\ \ ,\ \ N_f=2\ ,}
\nnn
1728J&=&\ds{\fc{(-\La_3^4-576\La_3^2m^2 + 3072\La_3m^3+256\La_3^2u-
4096u^2)^3}{2^{36}\Delta_3(u,m)}\ \ ,\ \ N_f=3\ ,}
\eea
with the discriminants:

\bea\label{deltas}
\ds{\Delta_0(u)}&=&\ds{2^{-6}\La^4(u^2-\La^4)^2}\nnn
\ds{\Delta_1(u,m)}&=&\ds{2^{-20}\La_1^6(27 \La_1^6+256 \La_1^3m^3-
288 \La_1^3mu-256m^2u^2+256u^3)}\nnn
\ds{\Delta_2(u,m)}&=&\ds{2^{-24}\La_2^4(\La_2^2+8m^2
- 8u)^2(\La_2^4 - 64\La_2^2m^2 + 16\La_2^2u + 64u^2)}\nnn
\ds{\Delta_3(u,m)}&=&\ds{2^{-32}\La_3^2(-3\La_3^3m -  
24\La_3^2m^2-2048\La_3m^3
-8\La_3^2u+768\La_3mu + 2048u^2)}\\
&\times&\ds{(-\La_3m - 8m^2 + 8u)^3\ ,}
\eea
we arrive at the periods $(\tilde\om_D,\tilde\om)$
given in \cite{klt,bf1,bf2} for $m=0$.
The above functions refer to the curves \cite{sw1,sw2}:

\be
y^2=(x^2-\La^4)(x-u)\ \ \ ,\ \ \ N_f=0\ ,
\ee
and 

\bea
y^2&=&\ds{x^2(x-u)+\fc{1}{4}m\La_1^3x-\fc{1}{64}\La_1^6\ \ \ ,\ \ \  
N_f=1}\nnn
y^2&=&\ds{\lf(x^2-\fc{1}{64}\La_2^4\ri)(x-u)+\fc{1}{4}m_1m_2\La_2^2x-
\fc{1}{64}(m_1^2+m_2^2)\La_2^4\ \ \ ,\ \ \ N_f=2}\nnn
y^2&=&\ds{x^2(x-u)-\fc{1}{64}\La_3^2(x-u)^2-\fc{1}{64}
(m_1^2+m_2^2+m_3^2)\La_3^2(x-u)\ +}\\
&+&\ds{\fc{1}{4}m_1 m_2 m_3\La_3 x-\fc{1}{64}(m_1^2m_2^2+m_2^2m_3^2+
m_1^2m_3^2)\La_3^2\ \ \ ,\ \ \ N_f=3\ .}
\eea
with $m_1=m_2=m_3=m$, respectively.
Choosing \req{mono0} and \req{mono1} for the periods of
the monopole patch the limits \req{limits} can be compared
with the monopole expansions in \cite{klt,ito,bf1,bf2} for $m=0$.

Since the coupling constant of the underlying theory

\be
\tau(J)=\fc{d^2\Fc}{da^2}=\fc{da_D}{da}=
\fc{da_D}{du}\slash\fc{da}{du}
\ee
is the quotient of two solutions of the hypergeometric equation
\req{unif},
it satisfies the {\em Schwarzian differential equation}
associated to eq. \req{unif}:

\be\label{schwarz}
\{\tau,J\}\equiv\fc{\tau'''}{\tau'}-\fc{3}{2}
\lf(\fc{\tau''}{\tau'}\ri)^2=
\fc{4}{9J^2}+\fc{3}{8(1-J)^2}+\fc{23}{72 J(1-J)}\ .
\ee
Its solution is known as the {\em Schwarzian triangle function}
$\om(J)\equiv s[\h,\fc{1}{3},0,J]$ which represents an infinite--valued
map
from the $J$--plane to the complex plane \cite{fricke}.
Notice that this provides an analytic expression for the
coupling constant $\tau$

\be
\tau(J)=\om[J]\equiv s\lf[\h,\fc{1}{3},0,J\ri]\ ,
\ee
modulo a subgroup of $SL(2,\Z)$ transformations. This function maps
the points $J=0, J=1$ and $J=\infty$ to the edges of a triangle
with angles $\pi/3, \pi/2$ and $0$ at the points $\tau=\rho,
i,i\infty$, respectively (with $\rho=e^{2\pi i/3}$). From

\be\label{modularinv}
\lf\{\fc{A\tau+B}{C\tau+D},J\ri\}=\{\tau,J\}\ \ ,\ \ AD-BC=1\ ,
\ee
it follows that any linear combination $\Om_D'=A\Om_D+B\Om$ and
$\Om'=C\Om_D+D\Om$ of the two solutions \req{solutions} also
satisfies
\req{schwarz}. It can be proven that all solutions of
\req{schwarz} are of this type.
To give an analytic expression for $\om(J)$ we have to select a particular
expansion in $J$.
When encircling the three singular points $J=\infty,0,1$ the two solutions
$\Om_0$ and $\Om_1$ have a certain monodromie behaviour which 
becomes clear when we look at the related points in the 
$\tau=\Om_0/\Om_1$--plane:

\begin{itemize}

\item $J=\infty$. The corresponding point $\tau=i\infty$ 
in the $\tau$--plane is invariant under $T: \tau\ra \tau+1$. 
Therefore at $J=\infty$:

$$\lf(\Om_0 \atop \Om_1\ri)\ra \lf( \ba{cc} 1 & 1\\ 0 &1 \ea\ri)
\lf(\Om_0 \atop \Om_1\ri)\ .$$

\item $J=1$. The corresponding point $\tau=i$ in the $\tau$--plane
is invariant under $S: \tau\ra -1/\tau$. Therefore at $J=1$:

$$\lf(\Om_0 \atop \Om_1\ri)\ra \lf( \ba{cc} 0 & 1\\ -1 &0 \ea\ri)
\lf(\Om_0 \atop \Om_1\ri)\ .$$

\item $J=0$. The corresponding point $\tau=\rho$ in the $\tau$--plane
is invariant under $T^{-1}S^{-1}: \tau\ra \tau+1$. Therefore at $J=0$:

$$\lf(\Om_0 \atop \Om_1\ri)\ra \lf( \ba{cc} -1 & -1\\ 1 &0 \ea\ri)
\lf(\Om_0 \atop \Om_1\ri)\ .$$

\end{itemize}
As expected, altogether the monodromies close: $T^{-1}S^{-1} ST=1$.
By these conditions the form of the solutions $\Om_0,\Om_1$ is fixed up to
normalization factors. To make this more precise we give two ratios for $\tau$,
one valid for $|J|<1$ and the other for $|J|>1$.
Since we know the monodromie behaviour
of the hypergeometric functions, we may write at $J=0$:

\bea\label{claudi}
\Om_0(J)-\rho^2\Om_1(J)&=&
\ds{A_0\F\lf[\fc{1}{12},\fc{1}{12},\fc{2}{3},J\ri]}\nnn
\Om_0(J)-\rho\Om_1(J)&=&
\ds{A_1\F\lf[\fc{5}{12},\fc{5}{12},\fc{4}{3},J\ri]\ .}
\eea
The combinations on the l.h.s. are eigenvectors with eigenvalues
$\rho,\rho^2$ under the monodromie action of $T^{-1}S^{-1}$.
The normalizations $A_0=\rho(1-\rho)$ and $A_1=-\rho\lambda\sqrt{3}$ with 

\be
\lambda=(2-\sqrt{3})
\lf[\fc{\Ga\lf(\fc{11}{12}\ri)}{\Ga\lf(\fc{7}{12}\ri)}\ri]^2
\fc{\Ga\lf(\fc{2}{3}\ri)}{\Ga\lf(\fc{4}{3}\ri)}
\ee
are determined when comparing the solutions \req{claudi} with the actual
integral \req{integral}. Therefore for $|J|<1$

\be\label{tau1}
\tau(J)=e^{\fc{2\pi i}{3}}\fc{\F\lf[\fc{1}{12},\fc{1}{12},
\fc{2}{3},J\ri]-\lambda e^{\pi i/3}
J^{\fc{1}{3}}\F\lf[\fc{5}{12},\fc{5}{12},\fc{4}{3},J\ri]}
{\F\lf[\fc{1}{12},\fc{1}{12},
\fc{2}{3},J\ri]-\lambda e^{-\pi i/3}
J^{\fc{1}{3}}\F\lf[\fc{5}{12},\fc{5}{12},\fc{4}{3},J\ri]}\ 
\ee
and for $|J|>1,\ |arg(1-J)|<\pi$:

\bea\label{tau2}
2\pi i\tau(J)=\ds{\fc{\Om_1(J)}{\Om_0(J)}}&=&\ds{
-\ln J-3 \ln 12 -\fc{\sum\limits_{n=1}
^\infty\fc{\lf(\fc{1}{12}\ri)_n\lf(\fc{5}{12}\ri)_n}{(n!)^2}h_nJ^{-n}}
{\F\lf[\fc{1}{12},\fc{5}{12},1,\fc{1}{J}\ri]}}\nnn
&=&\ds{-\ln J-3 \ln 12+\fc{31}{72}J^{-1}+\fc{13157}{82944}J^{-2}+\ldots\ .}
\eea
Indeed, these expressions imply $\tau(0)=e^{\fc{2\pi i}{3}},\  
\tau(1)=i$
and $\tau(\infty)=i\infty$. Altogether, we map the $J$--plane
onto half of the fundamental region of $SL(2,\Z)$, representing
a degenerate triangle.
Eqs. \req{tau1} and \req{tau2} allow us to determine the coupling
constant
$\tau$ in the whole $J$--plane and by \req{Js} we obtain its
full $u$ and $m$ dependence. Of course, as soon as we specify a
region
in the $u$--plane, the $SL(2,\Z)$ transformations are reduced to the
corresponding monodromy group of the relevant periods in this patch.
Inserting $J=j/1728$ in the above expressions and using
\cite{erdelyi3},
we immediately see that \req{tau1} and \req{tau2} become the coupling
constant of $N_f=4$.

\sect{Modular functions and periods}

Let us focus on the first solution of \req{solutions}, which
gives us the expression:

\be
\fc{da}{du}=\fc{\pi}{12}\sqrt{\fc{g_2}{g_3}}\  (J-1)^{\fc{1}{4}}
J^{-\fc{1}{4}}
\F\lf[\fc{1}{12},\fc{5}{12},1,\fc{1}{J}\ri]\ .
\ee
We apply the powerful identity \cite{robert}

\be\label{power}
\F\lf[\fc{1}{12},\fc{5}{12},1,\fc{1728}{j(\tau)}\ri]^4=E_4(\tau)
\ee
and

\be\label{eisen}
J=\fc{j(\tau)}{1728}=\fc{E_4(\tau)^3}{E_4(\tau)^3-E_6(\tau)^2}
\ee
to arrive at

\be\label{a0}
\fc{da}{du}=\fc{\pi}{12}\sqrt{\fc{g_2E_6(\tau)}{g_3E_4(\tau)}}\ .
\ee

For the massless $N_f=4$ case with the curve
$y^2=x^3-\fc{1}{4}g_2(\tau)xu^2-\fc{1}{4}g_3(\tau) u^3$ and 
$g_2(\tau)=\fc{4 \pi^4}{3} E_4(\tau),\ g_3(\tau)=\fc{8\pi^6}{27}
E_6(\tau),$
we just have to determine its $J$--function
\be
J=\fc{j(\tau)}{1728}\ \ ,\ \ N_f=4\ ,
\ee
to pass from \req{a0} to

\be
\fc{da}{du}=\h\fc{1}{\sqrt{2u}}\ .
\ee
Expression \req{a0} is a modular function of weight $+1$. Let us
remark that $g_2$ and $g_3$ are {\em invariants}
when considered as functions of $u$, since $u(\tau)$ is a modular
invariant
function in $\tau$.
The relation \req{power} indicates the existence of a deeper
connection of periods to modular functions.

Indeed, recently, it was shown in \cite{nahm} how to construct
the weight $-1$ modular function $c(\tau)=a_D-\tau a$
once its singularity structure in the $\tau$--moduli space is known.
Let us present some more connections for the $N_f=0$ case by writing
$J$ of \req{Js} completely in terms of modular functions.
Using the identity \cite{chand}

\be\label{lamb}
\fc{j(\tau)}{1728}=\fc{4}{27}
\fc{[1-\la(\tau)+\la^2(\tau)]^3}{\la^2(\tau)[1-\la(\tau)]^2}\ \ \ 
{\rm with\ \ } \la(\tau)=\fc{\th_1^4(\tau)}{\th_3^4(\tau)}
\ee
and eq. \req{Js}


\be\label{lamb1}
J=\fc{(3\La^4+u^2)^3}{27\La^4(u^2-\La^4)^2}=-\fc{1}{27}
\fc{(z-4)^3}{z^2}\ ,
\ee
we realize that the choice
$z=1-\fc{u^2}{\La^4}=4\fc{1}{\la}(1-\fc{1}{\la})$ matches eq.
\req{lamb}
with \req{lamb1}. This leads to

\be
\fc{u(\tau)}{\La^2}=-1+\fc{2}{\lambda(\tau)}\ ,
\ee
which indeed has the correct behaviour at the cusps
$\tau=0,1,i\infty$ corresponding to $u=1,-1,i\infty$, respectively.
This gives an invariant
expression for $u(\tau)$, since $\lambda(\tau)$
is a modular function of $\Ga(2)_\tau$ \cite{chand}, which is 
the monodromy group of the curve discussed in \cite{sw1}.

\ \\
$\underline{N_f=0:}$
\ \\
We want to discuss the curve $y^2=x^3-ux^2+\fc{1}{4}\La^4x$
discussed in \cite{sw2}. It has:

\bea\label{cindy0}
J&=&\ds{\fc{(-3\La^4+4u^2)^3}{27\La^8(u^2-\La^4)}}\nnn
g_2&=&\ds{\fc{1}{48}(4u^2-3\La^4)}\nnn
g_3&=&\ds{\fc{1}{1728}(8 u^3-9 \La^4 u)}\nnn
\ds{\triangle_0(u)}&=&\ds{\fc{\La^{12}}{4096}\lf(\fc{u^2}{\La^4}-1\ri)\  
.}
\eea
Its moduli space is
$\Ga^0(4)$ with the three cusps points $\tau=0,2,i \infty$.
The behaviour of $u(\tau)$ at the three cusps can be worked out:

\bea
\ds{\fc{u^4}{\La^8}}&=&\ds{\fc{27}{64} e^{-2\pi i\tau}
\ \ ,\ \ \tau\lra i\infty}\nnn
\ds{\fc{u}{\La^2}-1}&=&\ds{\fc{1}{54} e^{-2 \pi i \fc{1}{\tau}}
\ \ ,\ \ \tau\lra 0+i\epsilon}\nnn
\ds{\fc{u}{\La^2}+1}&=&\ds{-\fc{1}{54} e^{2 \pi i \fc{1}{2-\tau}}
\ \ ,\ \ \tau\lra 2+i\epsilon.}
\eea
These conditions may be deduced by writing $J(\tau)$ with a different
argument $J\lf(\fc{a\tau+b}{c\tau+d}\ri)$ depending on which power  
series one is interested in.
For example for the expansion around the monopole point one chooses
$J\lf(-\fc{1}{\tau}\ri)$, since $\tau_D=-\fc{1}{\tau}$ is the 
coupling of the dual theory which becomes weak $\tau_D\ra i\infty$
at the monopole point. Similarly, $J\lf(\fc{1}{\tau-2}\ri)$ at the dyon  
point
$\tau=2$.
Of course, since  
$J\lf(-\fc{1}{\tau}\ri)=J(\tau)=J\lf(\fc{1}{\tau-2}\ri)$ 
this makes no difference for the $J$--function. This is the effect
we have already encountered in the previous section, that $J=\infty$
holds for the monopole, the dyon and weak coupling point.
Eqs. \req{cindy0} are enough to determine the discriminant:

\be
\triangle_0(\tau)=\kappa_0\La^{12}
\fc{\eta^{24}\lf(\fc{\tau}{2}\ri)}{\eta^{24}(\tau)}\ \ \ ,\ \ \ 
\kappa_0=-2^{-18}\ 1728^{-2}\ ,
\ee
which is also invariant under $\Ga_0(2)_{\tilde\tau}$ with
$\tilde\tau=\fc{\tau}{2}$. If we were interested to match only the  
monopole behaviour at $\tau=0$ and the weak coupling behaviour at $\tau\ra  
i\infty$ we would obtain:

\be\label{ut1}
\fc{u(\tau)}{\La^2}=\fc{1}{2^3\ 1728}
\fc{\eta^8\lf(\fc{\tau}{4}\ri)}{\eta^8(\tau)}+1\ .
\ee
This expression has also been found in \cite{nahm}.
Similarly, when one only matches the dyon point $\tau=2$
and the weak coupling point $\tau=i\infty$, we get

\be\label{ut2}
\fc{u(\tau)}{\La^2}=-\fc{1}{2^3\ 1728}
\fc{\eta^{24}\lf(\fc{\tau}{2}\ri)}{\eta^{16}(\tau)\eta^8\lf(\fc{\tau}{4} 
\ri)}
-1\ .
\ee
These expressions are manifestly invariant only under $\Ga^0(4)\tau$.
Using

\bea\label{g2g3s}
\ds{g_{2,N_f}^3(\tau)}&=&\ds{\fc{1}{1728}\  
\fc{E_4^3(\tau)}{\eta^{24}(\tau)}
\triangle_{N_f}(\tau)}\nnn
\ds{g_{3,N_f}^2(\tau)}&=&\ds{\fc{1}{27\ 1728}\ 

\fc{E_6^2(\tau)}{\eta^{24}(\tau)}
\triangle_{N_f}(\tau)\ ,}
\eea
we determine:

\bea\label{cindy}
\ds{g_2(\tau)}&=&\ds{\La^4\fc{\kappa_0^{\fc{1}{3}}}{12}\ E_4(\tau)\ 

\fc{\eta^8\lf(\fc{\tau}{2}\ri)}{\eta^{16}(\tau)}}\nnn
\ds{g_3(\tau)}&=&\ds{\La^6\fc{\kappa_0^{\fc{1}{2}}}{216}\ E_6(\tau)\ 

\fc{\eta^{12}\lf(\fc{\tau}{2}\ri)}{\eta^{24}(\tau)}}\ .
\eea
Combining the two expressions for $g_2$ in eqs. \req{cindy0} and  
\req{cindy}
gives us the full $\tau$--dependence for $u(\tau)$, in contrast to
eqs. \req{ut1} and \req{ut2}:

\be
\fc{u^2(\tau)}{\La^4}=\kappa_0^{\fc{1}{3}}\ E_4(\tau)\ 
\fc{\eta^8\lf(\fc{\tau}{2}\ri)}{\eta^{16}(\tau)}+\fc{3}{4}\ .
\ee
Finally, with \req{a0} we obtain:

\be
\fc{da}{du}=\fc{\sqrt{2} \pi\kappa_0^{-\fc{1}{12}}}{4\La}
\ \fc{\eta^4(\tau)}{\eta^2\lf(\fc{\tau}{2}\ri)}\ .
\ee

\ \\
$\underline{N_f=2:}$ 

\ \\
The massless $N_f=2$ case\footnote{The massive case is more involved,
since the cusps move.} is described by the moduli space of
$\Ga(2)$ with the three cusps points $\tau=0,1,i \infty$.
At these points we obtain for $u(\tau)$

\bea
\ds{\fc{u^2}{\La^4_2}}&=&\ds{\fc{27}{64} e^{-2 \pi i \tau}\ \ ,
\ \ \tau\lra i\infty\ ,}\nnn
\ds{\lf(\fc{u}{\La_2^2}-u_{12}\ri)^2}&=&
\ds{\fc{1}{108}e^{-2\pi i \fc{1}{\tau}}
\ \ ,\ \ \tau\lra 0+i\epsilon\ \ \ ,\ \ u_{12} = \fc{1}{8}}\nnn
\ds{\lf(\fc{u}{\La_2^2}-u_{34}\ri)^2}&=&
\ds{{\fc{1}{108}e^{2\pi i \fc{1}{1-\tau}}\ \ ,
\ \ \tau\lra 1+i\epsilon\ ,\ \ u_{34} = -\fc{1}{8}}\ ,}
\eea
from which we get

\be
\triangle_2(\tau)=\kappa_2\La_2^{12}
\fc{\eta^{48}(\tau)}{\eta^{48}(2\tau)}\ \ \ ,\ \ \ 
\kappa_2=2^{-36}\ 1728^{-2}\ .
\ee
It is interesting that this expression has an even bigger
invariance, namely $\Ga_0(2)_\tau$, as it would be dictated by the 
monodromy group $\Ga(2)_\tau$.
Using \req{g2g3s} we obtain for  $g_2$ and $g_3$

\bea
\ds{g_2(\tau)}&=&\ds{\La_2^4\ \fc{\kappa_2^{\fc{1}{3}}}{12}
\ E_4(\tau)\fc{\eta^8(\tau)}{\eta^{16}(2\tau)}\ ,}\nnn
\ds{g_3(\tau)}&=&\ds{\La_2^6\ \fc{\kappa_2^{\fc{1}{2}}}{216}
\ E_6(\tau)\fc{\eta^{12}(\tau)}{\eta^{24}(2\tau)}\ .}
\eea
From the explicit form of $g_2=\fc{1}{768}(3\La_2^4+64 u^2)$
we are able to extract $u(\tau)$.
With \req{a0} we arrive at:

\be
\fc{da}{du}=\fc{\sqrt{2}\pi\kappa_2^{-\fc{1}{12}}}{4\La_2}
\fc{\eta^4(2\tau)}{\eta^2(\tau)}\ .
\ee

\ \\
$\underline{N_f=3:}$
\ \\
The massless $N_f=3$ case is described by the moduli space of
$\Ga_0(4)$ with the three cusps points $\tau=0,\fc{1}{2},i\infty$.
From the behaviour at the three cusps

\bea
\ds{\fc{u}{\La^2_3}}&=&\ds{-\fc{27}{64} e^{-2 \pi i \tau}\ \ ,
\ \ \tau\lra i\infty\ ,}\nnn
\ds{\lf(\fc{u}{\La_3^2}-u_{1234}\ri)^4}&=&
\ds{\fc{1}{2^{22}3^3}e^{-2\pi i \fc{1}{\tau}}
\ \ ,\ \ \tau\lra 0+i\epsilon\ \ \ , \ \ u_{1234} = 0 \,}\nnn
\ds{\fc{u}{\La_3^2}-u_5}&=&\ds{\fc{-1}{2^{10}3^3}e^{-2\pi i  
\fc{\tau}{2\tau-1}}
\ \ ,\ \ \tau\lra
\fc{1}{2}+i\epsilon\ \ , \ \ u_5 = \fc{1}{256} \ .}
\eea
we are able to determine the discriminant

\be
\triangle_3(\tau)=\kappa_3\La_3^{12}
\fc{\eta^{24}(\tau)\eta^{24}(2\tau)}{\eta^{48}(4\tau)}\ \ \ ,\ \ \ 
\kappa_3:=-2^{-132}\ 1728^{-5}\ .
\ee
which is the modular invariant expression for $\triangle_3(u,m)$
in \req{deltas}. This function must be a modular invariant
under $\Ga_0(4)_\tau$.
In addition, it is also invariant under $\Ga(2)_{\tilde \tau}$
for $\tilde\tau=2\tau$ arising from $\Ga_0(4) \simeq \Ga(2)$.
In other words, this just reflects the fact that we can also describe  
the monodromies of the $N_f=3$ case in the $\tilde\tau$
moduli space.
Let us also mention the identity \cite{koblitz}

\be
\triangle_3(\tau)=\kappa_3 \La_3^{12}
e^{-\fc{\pi i}{3}}\fc{\eta^{32}(\tau)}{\eta^{32}(4\tau)}\ 
\fc{\eta^8(\tau+\fc{1}{2})}{\eta^8(4\tau)}\ ,
\ee
to realize that our discriminant is the product of two single functions
describing the correct behaviour at $\tau=0$ and $\tau=\h$,
respectively.
Furthermore, using \req{g2g3s}, we may evaluate:

\bea
\ds{g_2(\tau)}&=&\ds{\La_3^4\ \fc{\kappa_3^{\fc{1}{3}}}{12}
\ E_4(\tau)\ \fc{\eta^8(2\tau)}{\eta^{16}(4\tau)}\ ,}\nnn
\ds{g_3(\tau)}&=&\ds{\La_3^6\ \fc{\kappa_3^{\fc{1}{2}}}{216}
\ E_6(\tau)\ \fc{\eta^{12}(2\tau)}{\eta^{24}(4\tau)}\ .}
\eea
With $g_2=\fc{1}{49152}(\La_3^4-256\La_3^2 u+4096 u^2)$ we are able to
deduce $u(\tau)$.
Finally, from \req{a0} we obtain:

\be
\fc{da}{du}=\fc{\sqrt{2}\pi\kappa_3^{-\fc{1}{12}}}{4\La_3}
\ \fc{\eta^4(4\tau)}{\eta^2(2\tau)}\ .
\ee

Let us now come to a case with $m\neq 0$.

\ \\
$\underline{N_f=3\ ,\ m=\fc{\La_3}{8}}:$
\ \\
This case corresponds to a superconformal point (see sect. 4 for  
discussions).
Relating the $J$--function (3.3) with its explicit form (2.17) gives
two branches for $u(\tau)$:

\be\label{SCFT3}
\fc{u(\tau)}{\La_3^2}=
\fc{\pm E_4(\tau)^3\lf[-23+27\sqrt{\fc{E_6(\tau)^2}{E_4(\tau)^3}}\ri]
-4E_6(\tau)^2}{128[E_4(\tau)^3-E_6(\tau)^2]}
\ .
\ee
Similar expressions may be found for $g_2(\tau),\ g_3(\tau)$ to 
derive from (3.4):

\be
\fc{da}{du}=\fc{4 \sqrt 3 \pi}{9\La_3}E_4(\tau)^{1/4}
\sqrt{\pm 1+
\fc{E_6(\tau)}{E_4(\tau)^{3/2}}}\ .
\ee

To summarize, we have explicitly demonstrated for $N_f=0,2,3$ that one  
may
express $\fc{da}{du}$ and therefore also $\fc{da_D}{du}$
by a modular function $c(\tau)$ of weight $-1$:

\bea\label{das}
\ds{\fc{da}{du}}&=&\ds{c^{(-1)}(\tau)}\nnn
\ds{\fc{da_D}{du}}&=&\ds{\tau\ c^{(-1)}(\tau)\ .}\nnn
\eea
However, from these expressions we are able to extract $a_D$ and $a$
up to an integration constant containing possible residua in the  
massive
case. From \cite{sty} we have:

\be
a_D-\tau a=\lf(\fc{da}{du}\ri)^{-1}\lf[\fc{i}{2\pi}(4-N_f)-
\sum_i m_i\int_{x_i}^\infty \fc{dx}{y}\ri]\ .
\ee
The $x_i$ are the locations of the residua on the hyperelliptic
curve.
Using \req{das} we obtain

\be
a_D-\tau a=c(\tau)\lf[\fc{i}{2\pi}(4-N_f)-
\sum_i m_i\int_{x_i}^\infty \fc{dx}{y}\ri]\ ,
\ee
from which we immediately get expressions for the periods $a_D$ and
$a$ in the case $m=0$:

\bea
a&=&\ds{-\fc{i}{2\pi}(4-N_f)\ \fc{dc}{d\tau}}\nnn
a_D&=&\ds{\fc{i}{2\pi}(4-N_f)\lf[c-\tau\fc{dc}{d\tau}\ri]\ .}
\eea

\sect{$J$ Invariants and monodromies}

In this section we will use the explicit form of the $J$--functions
\req{Js} of
the Seiberg--Witten curves in terms of the $u$, $m$ and $\La$,  to
obtain the
behaviour near the singularities of the moduli space. We are mainly
interested
in the pole structure of the $J$--function because from that we can
derive
the monodromies (up to conjugation). Near the cusp $\tau = i \infty$
the
$J$--invariant takes the following form

\be J(\tau) \sim q^{-1} = e^{-2 \pi i \tau}\ , \ee
as can be seen from \req{tau2}.
A loop in $J$--space $J \rightarrow J e^{2 \pi i}$ corresponds to the
shift
$\tau \rightarrow \tau - 1$ or in other words the associated
monodromy is
$T^{-1}$. Furthermore we will find points in the moduli space of the
massive theories where $J = 0$ or $J = 1$. We will show that these
points correspond to
superconformal points, which have been studied in the literature
\cite{apsw}.
We find that the coupling constant at these points is $\rho$ or $i$
and
the associated monodromy is conjugate to $(S T)^{-1}$ or $S^{-1}$.

Let us begin with the case $N_f = 0$. The $J$--invariant \req{cindy0} 
has three poles at $u= \pm \La^2$ and $u = \infty$, which 
correspond to the two strong
coupling
singularities and the weak coupling singularity. At these poles $J$
shows the
following behaviour:

\bea\label{jnf0}
J & = & \ds{\pm \frac{\La^2}{54} (u \mp \La^2)^{-1} \ \ , \ \ u = \pm
\La^2} \nnn
J & = & \ds{\frac{64}{27 \La^4} u^4 \ \ , \ \ u = \infty \ .}
\eea
If $u$ moves around the singularity at infinity
$u \rightarrow u e^{2 \pi i}$,
$J$ loops four times, which means that the weak coupling monodromy is
conjugate to $T^{-4}$. Along the same line we conclude that the
monodromies at the
strong coupling singularities are conjugate to $T$, in agreement with
\cite{sw2}. At the monopole point
$u = \La^2$ we have $\tau = 0$, which corresponds to strong coupling,
and
we have to perform a duality transformation, which takes $\tau$ to
infinity and
introduce a dual coupling $\tau_D = -1/\tau$ with
$\tau_D \rightarrow i \infty$ for $\tau \rightarrow 0$. In a more
physical language: we go from a strongly coupled description of the  
theory to a weakly
coupled description with coupling constant $\tau_D$. Therefore we
find

\be J = e^{-2 \pi i \tau_D} = e^{2 \pi i/ \tau} \sim c_+ (u -
\La^2)^{-1} \ee
\be \Rightarrow u = \La^2 + c_+ e^{- 2 \pi i/ \tau}\ , \ee
and similarly at the dyon point $u = - \La^2$, where $\tau = 2$, we
a duality transformation $\tau_{D'} = 1/(2 - \tau)$ and find:

\be \ds{u = -\La^2 + c_- e^\frac{2 \pi i}{2 - \tau} \ .} \ee
The behaviour at the singularities fixes $u$ uniquely to be a
$\Ga^0(4)$ modular function. Since the poles of $J$ correspond to
$\tau$
being either $\infty$ or a rational number, 
we can always transform this point to $i \infty$
by an $SL(2,\Z)$ duality transformation. Furthermore these points are
always
cusps of infinite order of the corresponding fundamental domain.

We will repeat the same arguments for the case $N_f = 1$, where we
have three
strong coupling singularities and a weak coupling singularity for
generic
values of $m$. By $u_i , i =  1, \ldots,3$ we denote the three zeros
of the
$N_f = 1$ discriminant \req{deltas}. The behaviour at the
singularities is

\bea\label{jnf1}
J & = & c_i (u - u_i)^{-1} \ \ , \ \ u = u_i       \nnn
J & = & \ds{ \frac{-64}{27 \La_1^6} u^3 \ \ , \ \ u = \infty\ ,}
\eea
where $c_i , i = 1, \ldots, 3$ are factors depending on $m$ and
$\La_1$ to be determined from $J$ in \req{Js}.
The monodromies at the three dyon points are conjugate to $T$ and the
weak
coupling monodromy is conjugate to $T^{-3}$, as expected \cite{sw2}.
As was
shown in \cite{apsw}
the moduli space of the $N_f = 1$ theory contains a superconformal
point
at $u = 3/4 \La_1^2 , m = 3/4 \La_1$ where two mutually non--local
particles become massless. In the notation of \cite{apsw}
this corresponds to a $(1,1)$ superconformal point, which is
equivalent
to the
superconformal point of the $SU(3)$ $N = 2$ SYM theory studied in
\cite{ad}.
At an $(n,1)$ superconformal point there are $n$ mutually local
massless particles
 together with one particle that is non--local with respect to them.
Indeed if we set $m = 3/4 \La_1$ the discriminant \req{deltas}
develops a
simple zero at $u = -15/16 \La_1^2$ and a double zero at $u = 3/4
\La_1^2$.
The $J$--invariant takes the following form:

\be J_{(1,1)}(u) = \frac{4 (3 \La_1^2 - 4 u)(3 \La_1^2 + 4 u)}{27
\La_1^6
(15 \La_1^2 + 16 u)} \ . \ee
For the simple zero $J$ has a simple pole and a monodromy conjugate
to $T$, whereas for $u = 3/4 \La_1^2$ we find that $J = 0$. Locally $J
\sim
-128 (u - 3/4 \La_1^2)/(27 \La_1^2)$. This means that at the
superconformal
point $\tau_{eff} = \rho = (1 + i \sqrt{3})/2$, since $J(\rho) = 0$ and
the
monodromy is conjugate to $(S T)^{-1}$, which is of order 3. This  
implies that the
coupling constant
has no log-dependence at the superconformal point. We expect to 
happen this precisely for a conformal theory. Away from these
special points all singularities are cusps of infinite
order, which is responsible for the log--dependence of the effective
coupling constant.
This value of the coupling was also found for the superconformal
point of \cite{ad}.
The order 3 is natural in the sense that there are in fact three
possibilities that two of three strong coupling singularities can
collide.

For the $N_f = 2$ case we will restrict ourselves to the case
$m_1 = m_2 = m$
to keep formulas simpler. The $N_f = 2$ discriminant \req{deltas} has
two
simple zeros at $u_{1,2} =
- \La_2^2/8 \mp \La_2 m$ and one double zero at $u_3 = \La_2^2/8 +
m^2$. The $J$--function near the singularities behaves as follows:

\bea\label{jnf2}
J & = & \ds{\mp \frac{\La_2(\pm \La_2 + 2 m)^2}{216 m} (u - u_i)^{-1} \  
\
, \ \ u = u_{1,2} }      \nnn
J & = & \ds{ \frac{(-\La_2^2 + 4 m^2)^4}{108 \La_2^4} (u - u_3)^{-2} \  
\ ,
\ \ u = u_3 \ }      \nnn
J & = & \ds{ \frac{64}{27 \La_2^4} u^2 \ \ , \ \ u = \infty \ . }
\eea
At $u = u_{1,2}$ we have $\tau = 1$, which is one of the three cusps
of
the fundamental domain of the massless $N_f = 2$ theory $\Ga(2)$,
which has cusps
of infinite order at $\tau = 0 , 1$ and $i \infty$. We find two
monodromies conjugate to $T$,
which collide for $m = 0$ where they produce a monodromy conjugate to
$T^2$.
They correspond to two mutually local particles and therefore they
cannot
generate a superconformal point at $m = 0$.
On the other hand there is a monodromy conjugate to $T^2$ at $u =
u_3$ and
$\tau = 0$, which splits into
two singularities with monodromies conjugate to $T$ for generic
values of
$m_1$ and $m_2$.
There exist two (2,1) superconformal points $m = \pm \La_2/2$,
$u = 3/8 \La_2^2$ and two more if we choose $m_1 = -m_2$. Near this
point
$J$ takes the following form:

\be J_{(2,1)} = \frac{(3 \La_2^2 + 8 u)^3}{27 \La_2^4 (5 \La_2^2 + 8
u)}
\ .\ee
At $u = 3/8 \La_2^2$ we have $J = 1$, which implies that $\tau_{eff} =
i$ at
the superconformal point and the respective monodromy is conjugate to
$S^{-1}$.
Notice, at this point it is not $u_1$ and $u_2$ which collide 
(they correspond to mutually
local dyons which contribute to the same cusp), but $u_2$ and $u_3$ 
or $u_1$ and $u_3$ for $m=\pm\La_2/2$, respectively,
where two
mutually local and one mutually non--local particles become massless.

Finally, we consider the $N_f = 3$ case. Again, we choose all
masses
to be equal $m_1 = m_2 = m_3 = m$. In the massless case there are two
strong
coupling monodromies, one at $u_1 = 0$, which corresponds to four
monopoles and one
at $u_2 = \La_3^2/256$ with a dyon of quantum numbers $(2,1)$. The
leading terms
of the $J$--invariant at the singularities are for $m = 0$:

\bea\label{jnf3}
J & = & \ds{ \frac{\La_3^8}{2^{22} 3^{3}} u^{-4} \ \ , \ \ u = u_1}\nnn
J & = & \ds{-\frac{\La_3^2}{2^{10} 3^{3}} (u - u_2)^{-1} \ \ , \ \ u =  
u_2}
\nnn
J & = & \ds{-\frac{64}{27 \La_2^4} u \ \ , \ \ u = \infty \ .}
\eea
It is easy to see that this is in agreement with the form of the
monodromies, 
i.e. they are conjugate to $T^4$, $T$ and $T^{-4}$, respectively. If
we turn
on the mass $m$ the global flavour symmetry $SU(4)$ is broken to
$SU(3) \times
U(1)$ and the four--fold singularity at $u = 0$ splits up into a
simple
singularity and a threefold singularity with monodromies conjugate to
$T$ and
$T^3$. As pointed out in \cite{apsw} there are several possibilities
to get
superconformal points of type (1,1) and (2,1) for varying masses,  but
we only
treat the special
points of type (3,1). This point occurs at $m = \La_3/8$, 
 where the $(2,1)$ dyon singularity coincides with
the triple singularity $u = \La_3^2/32$ and there is another dyon 
singularity at $u = - 19 \La_3^2/256$. For this special value of $m$ the 
$J$--function takes the form:

\be J_{(3,1)}(u) = \frac{-16 (32 u - \La_3^2 )^2}
{27 \La_3^2(19 \La_3^2 + 256 u)} \ .\ee
At the superconformal point $J$ vanishes, which means that
$\tau_{eff} = \rho$. With $E_4(\rho)=0$, we recover that
very easily from \req{SCFT3}.
The monodromy is conjugate to $(S T)^{-2}$ and of order 3.


\ \\ \\
{\em {\mbox{\boldmath $Acknowledgement:\ $}}}
We are very grateful to A. Klemm, W. Lerche, P. Mayr and S. Theisen
for helpful discussions.
We also thank J.--P. Derendinger for providing excellent
working conditions in Neuch\^atel.
The research of A. B. has been supported in part by the
Bundesministerium f\"ur
Wissenschaft, Verkehr und Kunst.


\end{document}